\documentclass[showpacs,pra,twocolumn,10pt,superscriptaddress]{revtex4-2}
\usepackage{epsfig,epstopdf,graphicx,amssymb,amsmath,color,bm}
\usepackage[colorlinks,urlcolor=blue,linkcolor=blue,citecolor=blue]{hyperref}  
\usepackage{amsmath} % 引入 amsmath 包以支持数学公式  
\usepackage{amsfonts} % 可选，用于字体支持  
\usepackage{amssymb} % 可选，用于额外的数学符号支持  
\usepackage{bibunits}
\begin{document}

\title{Pulse-Controlled Topologically Protected Quantum Batteries}
\author{Jin Yang}
\affiliation{College of Physics and Electronic Science, Hubei Normal University, Huangshi 435002, China}

\author{Biao Xiong}
\thanks{bx\_hbnu@163.com}
\affiliation{College of Physics and Electronic Science, Hubei Normal University, Huangshi 435002, China}

\author{Jibing Liu}
\affiliation{Hubei Engineering Institute, Huangshi 435006, China}
\affiliation{College of Physics and Electronic Science, Hubei Normal University, Huangshi 435002, China}

\author{Houguo Yu}
\affiliation{College of Physics and Electronic Science, Hubei Normal University, Huangshi 435002, China}

\author{Dehua Liu}
\affiliation{College of Physics and Electronic Science, Hubei Normal University, Huangshi 435002, China}

\author{Feng Mei}
\thanks{meifeng@sxu.edu.cn}
\affiliation{State Key Laboratory of Quantum Optics and Quantum Optics Devices, Institute of Laser Spectroscopy, Shanxi University, Taiyuan, Shanxi 030006, China}
\affiliation{Collaborative Innovation Center of Extreme Optics, Shanxi University, Taiyuan, Shanxi 030006, China}

\author{Chuanjia Shan}
\thanks{cjshan@hbnu.edu.cn}
\affiliation{College of Physics and Electronic Science, Hubei Normal University, Huangshi 435002, China}

\begin{abstract}
Quantum batteries have emerged as a promising new generation of energy-storage devices for powering quantum technologies. Long-distance charging is particularly attractive because it minimizes interference between the charger and the battery, thereby attracting considerable interest. Here, we propose a topologically protected long-distance charging protocol for quantum batteries based on a pulse-controlled superconducting qubit chain. By dynamically modulating the pulse-mediated couplings, we realize topologically protected energy transfer from the charger to the battery. We show that the charging process is free of energy backflow and remains robust against imperfections in pulse control. Moreover, the energy stored in the battery at the target time is fully extractable, and the protocol remains effective for relatively large system sizes. To further accelerate charging, we optimize the pulse shape and elucidate the underlying physical mechanism. Our pulse-controlled topological quantum battery protocol provides a versatile framework for implementing long-distance topological charging and establishes a theoretical foundation for designing optimal-control strategies to enhance quantum battery performance.
\end{abstract}

%\pacs{42.50.Wk, 07.10.Cm, 03.65.Yz, 42.50.Lc}

%\begin{spacing}{2.0}

\maketitle

\textit{Introduction---}Quantum batteries (QBs), as a class of innovative energy storage devices, harness quantum resources such as coherence and entanglement to achieve potential advantages in storage capacity and charging power\cite{RevModPhys.96.031001,PhysRevLett.129.130602,PhysRevB.104.245418,PhysRevE.102.052109,PhysRevLett.111.240401,PhysRevLett.128.140501,PhysRevA.110.022433}. In contrast to conventional chemical batteries, QBs can be seamlessly incorporated into quantum architectures via dipole-dipole couplings, photon-mediated interactions, or other forms of light-matter interaction, thereby offering a flexible energy source for quantum technologies~\cite{l39v-jwwz}. Over the past decade, significant theoretical breakthroughs have been made in the development of QBs, including the introduction of rigorous performance metrics such as stored energy and maximum extractable work (ergotropy) \cite{AEAllahverdyan_2004}. Various mechanisms have been proposed to enhance QB performance, encompassing the adiabatic evolution protocol\cite{PhysRevE.100.032107}, the indirect charging scheme mediated by intermediate states~\cite{PhysRevB.99.035421,PhysRevResearch.4.033216},  the combined effects of beam-splitter and parametric amplification~\cite{liu2026dissipative}, the collective charging effect based on Dicke or Tavis-Cummings models\cite{PhysRevA.109.022210,PhysRevA.109.012204,PhysRevB.98.205423}, coherent ergotropy ratio optimization~\cite{d9k1-75d4,060602-2026-0106}, and the Mpemba effect~\cite{PhysRevLett.134.220402,5xrr-x2rm}. In parallel, experimental efforts have realized fundamental proof-of-principle validations of QBs across multiple platforms, most notably in superconducting circuits and trapped-ion systems \cite{doi:10.1126/sciadv.abk3160,PhysRevA.106.042601,Hu_2022,sp5l-c6m8}.

However, in the practical implementation of QBs, the performance is inevitably compromised by several critical factors, such as energy backflow from the battery to the charger, environment-induced decoherence, and perturbations due to random system noise. Consequently, a central challenge in transitioning QBs from proof-of-principle to practical applications is to enhance performance under the influence of these factors. To address the issue of energy backflow, recent research has introduced a nonreciprocal quantum battery scheme, which fundamentally prevents backflow losses by achieving entirely unidirectional energy transfer \cite{PhysRevLett.132.210402,p93y-jflt,fn1b-2m9g,67wh-1fxv,kh36-7z76,liu2026chiralquantumbatteries}. To mitigate the adverse effects of environmental-induced decoherence, various reservoir engineering strategies have been proposed. These include continuous energy supply through structured reservoirs with bound states \cite{PhysRevA.102.060201,PhysRevLett.132.090401}, utilization of the non-Markovian effect to significantly improve performance \cite{Kamin_2020, Li:22,PhysRevA.109.012224,PhysRevA.104.032207,PhysRevA.106.012425,PhysRevA.102.052223}, interference engineering in giant-atom system~\cite{43n6-rnj3} and reservoir-assisted charging with quantum feedback control~\cite{vqnk-kzqg,2025arXiv251107134Y} or squeezing~\cite{b5lx-j66h}. 

Moreover, time-dependent quantum control has been employed in various QB protocols for different purposes~\cite{Hu_2022,PhysRevA.107.032218,Evangelakos_2025,j35s-xv5k}, such as accelerating evolution to suppress decoherence and preventing energy backflow during charging. However, the presence of noise in classical control fields inevitably degrades QB performance. This degradation is especially severe for large-scale remote charging of many-body systems, as the required time-dependent control involves a large number of control fields, each subject to fluctuations. How one can achieve robust quantum battery charging against classical control field noise in remote many-body systems remains an open challenge.

Topology, as a fundamental concept in many-body systems, provides a powerful framework for understanding and engineering robust quantum phenomena, benefiting from the protection of topologically nontrivial structures~\cite{ RevModPhys.91.015006,Luo2022-ov,6-20231850,qi2020progress}. Recent advances in topological photonics have opened promising avenues for enhancing the robustness of quantum optical effects. In the context of quantum batteries, the concept of topological quantum batteries have been proposed~\cite{PhysRevLett.134.180401}, where bound states in the nontrivial topological phase enable near-perfect energy transfer, and the presence of a dark state together with a topologically robust dressed bound state renders the battery immune to single-sublattice dissipation. Separately, a $\mathcal{PT}$-symmetric SSH quantum battery has been proposed~\cite{4klp-kw27}, in which topology and non-Hermiticity cooperate to yield enhanced charging dynamics. Both schemes rely on static control fields with no extra control, and thus inevitably suffer from energy backflow during the charging process. Motivated by these pioneering works, we introduce time-dependent control into topological quantum batteries to suppress energy backflow and enhance robustness against noise.

Superconducting quantum circuits, with chip-level programmability, fast all-electrical controls, and scalable architecture, provide a versatile solid-state platform for exploring topological effects~\cite{PhysRevLett.123.080501,doi:10.1126/science.adp6802,Xu_2023,Jin2025-sq,Zhang2022-sg,PhysRevLett.132.020601,PhysRevLett.134.070601,Li2020-nh,PhysRevLett.131.080401,PhysRevLett.133.140402,PhysRevLett.125.160503,PhysRevLett.132.036603}. Motivated by this, we propose a topological SSH-type superconducting qubit chain as a quantum battery model, where the left and right edge states serve as the charger and the battery, respectively. By designing the couplings as cosine-type pulses, we realize a topologically protected long-range charging channel that exhibits no energy backflow and is robust against noise in the control fields. Furthermore, we optimize the control protocol by introducing tan-type pulses, which significantly accelerate the topologically protected charging process. Our results demonstrate a practical and flexible route for achieving high-performance, noise-resistant remote quantum battery charging in scalable solid-state architectures.

\begin{figure}[t]
\centering
\includegraphics[width=0.48\textwidth]{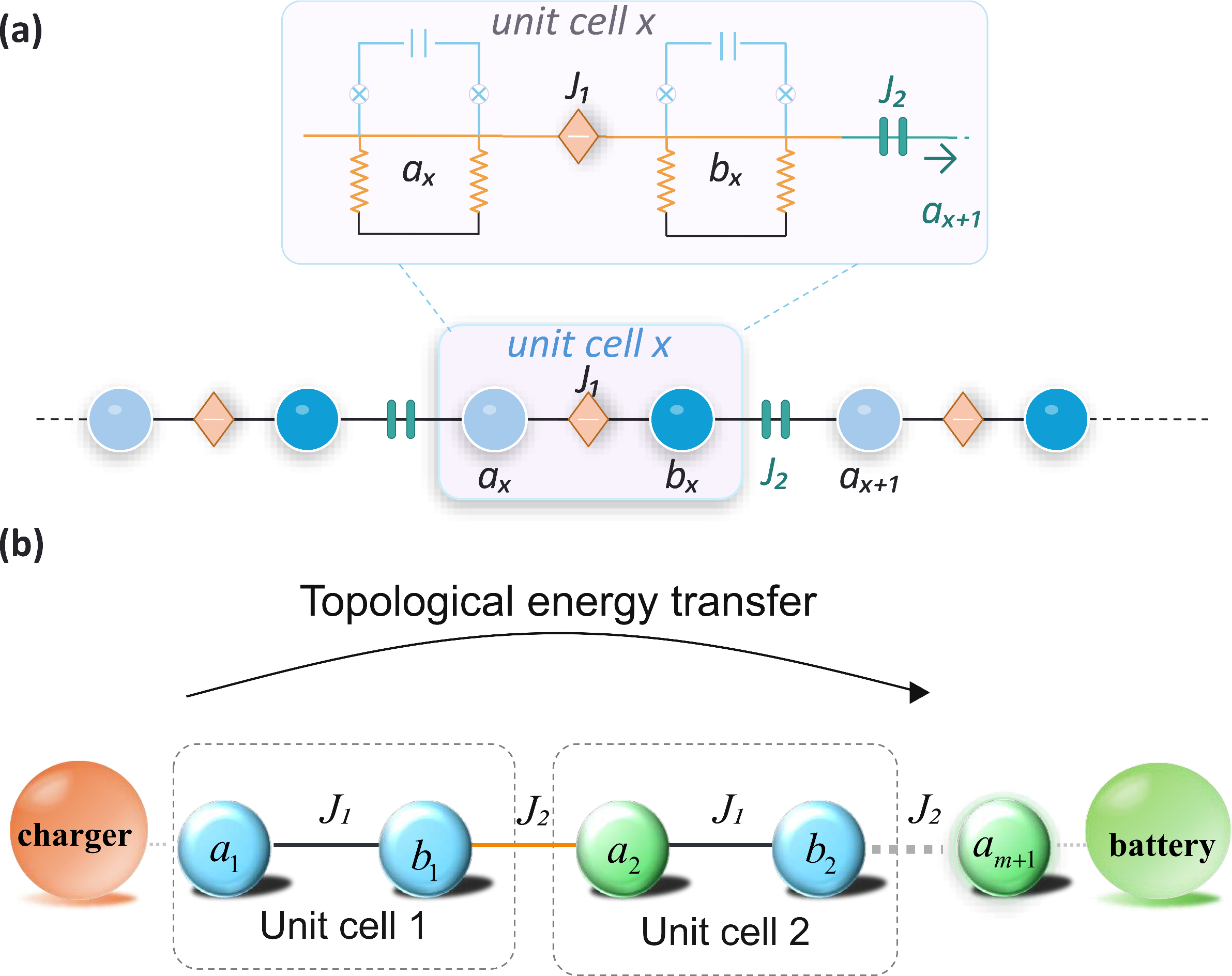}
\caption{(a) Schematic of the SSH-type qubit chain. Each unit cell contains two qubits $a_x$ and $b_x$, with intra-cell coupling $J_1(t)$ and inter-cell coupling $J_2(t)$. (b) Schematic of the topological quantum battery. The leftmost qubit $a_1$ serves as the charger, the rightmost qubit $a_{m+1}$ as the battery, and the intermediate qubits form a topologically protected channel for energy transfer.}
\label{fig:device}
\end{figure}

\textit{Model and Hamiltonian---}We consider a one-dimensional SSH-type qubit chain consisting of $N$ qubits, as illustrated in Fig.~\ref{fig:device}, with each unit cell containing two qubits labeled as $a_x$ and $b_x$ ($x=1,2,\ldots,m$). The Hamiltonian of the system is given by
\begin{align}
\hat{H}(t)=\sum_{x=1}^{m}\left[J_1(t)\hat{\sigma}_{a_x}^{+}\hat{\sigma}_{b_x}^{-}+J_2(t)\hat{\sigma}_{b_x}^{+}\hat{\sigma}_{a_{x+1}}^{-}+\mathrm{H.c.}\right],
\label{eq:H}
\end{align}
where $\hat{\sigma}_{a_x}^{+}=|e\rangle_{a_x}\langle g|$ and $\hat{\sigma}_{a_x}^{-}=|g\rangle_{a_x}\langle e|$ are the raising and lowering operators for the qubit at site $a_x$, and similarly for $b_x$. The SSH model presented here applies generically to a wide range of qubit platforms. For a concrete implementation, we turn to the superconducting Xmon architecture\cite{PhysRevA.98.012331}, illustrated in Fig.~\ref{fig:device}(a), in which the coupler current provides continuous control over the nearest-neighbor coupling strength $J_{j}(t)$ $(j=1,2)$~\cite{PhysRevLett.113.220502}.

In our quantum battery protocal, the leftmost qubit $a_1$ serves as the quantum charger, while the rightmost qubit $a_{m+1}$ functions as the quantum battery, as shown in Fig.~\ref{fig:device}(b). The intermediate qubits form a topologically protected channel for energy transfer. We assume that the qubit acting as the charger is initialized in the excited state, while all other qubits are prepared in their ground states. Owing to energy conservation of the SSH model, the basis vectors $\{|e,g,g,\ldots\rangle, |g,e,g,\ldots\rangle, \ldots\}$ form a complete set. In this basis, the Hamiltonian takes the matrix form $H_M$, with elements $\left(H_{M}\right)_{i,j} = J_1(t)$ for $j = i \pm 1$ with $i$ odd, $J_2(t)$ for $j = i \pm 1$ with $i$ even, and $0$ otherwise. Further calculation reveals that the eigenvalues of $H_M$ are $E_{-}^{(m)}, \ldots, E^{(0)}, \ldots, E_{+}^{(m)}$ in ascending order, with corresponding eigenstates $|\psi\rangle_{-}^{(m)}, \ldots, |\psi\rangle^{(0)}, \ldots, |\psi\rangle_{+}^{(m)}$. The explicit expressions for these eigenvalues and eigenstates are given in the Supplemental Material. Here, we focus on the central zero eigenvalue $E^{(0)} = 0$, whose eigenstate is
\begin{equation}
|\psi\rangle^{(0)}=\frac{1}{\sqrt{\eta_0}}\left[(-\nu)^0, 0,(-\nu)^1, 0, \ldots, (-\nu)^{m}\right]^T,
\label{eq:eigenstate0}
\end{equation}
with $\nu = J_1/J_2$. It can be seen that the zero-energy eigenstate $|\psi\rangle^{(0)} \approx (1,0,0,\ldots)^T=|e,g,g,\ldots\rangle$ for $|\nu| \ll 1$, while $|\psi\rangle^{(0)} \approx (0,\ldots,0,1)^T=|g,\ldots,g,e\rangle$ for $|\nu| \gg 1$. This observation allows us to tune the value of $\nu=J_1/J_2$ such that $|\psi\rangle^{(0)}$ can be switched from the left edge state $|e,g,g,\ldots\rangle$ to the right edge state  $|g,\ldots,g,e\rangle$, thereby realizing a energy transition from the left-localized qubit to the right-localized qubit. In fact, there are many choices of $J_1(t)$ and $J_2(t)$ that satisfy this condition. Without loss of generality, we assume $J_1(t)$ and $J_2(t)$ as follows:
\begin{align}
J_1(t) &= J_0\left[1 - f(t) \right], \quad
J_2(t) = -J_0\left[1 + f(t)\right],
\label{eq:J_pulses}
\end{align}
where $J_0$ is the coupling constant, $f(t)$ satisfies $f(0)=1$ and $f(t_f)=-1$ with $t_f$ the target time. Clearly, $|\nu(0)| = 0$ and $|\nu(t_f)| = \infty$, which satisfy the above condition and thus enable the energy transfer from the left edge state to the right edge state. In the following, we utilize these topological edge states as energy carriers to realize directional energy transfer from the charger to the battery.

\begin{figure}[t]
\centering
\includegraphics[width=0.48\textwidth]{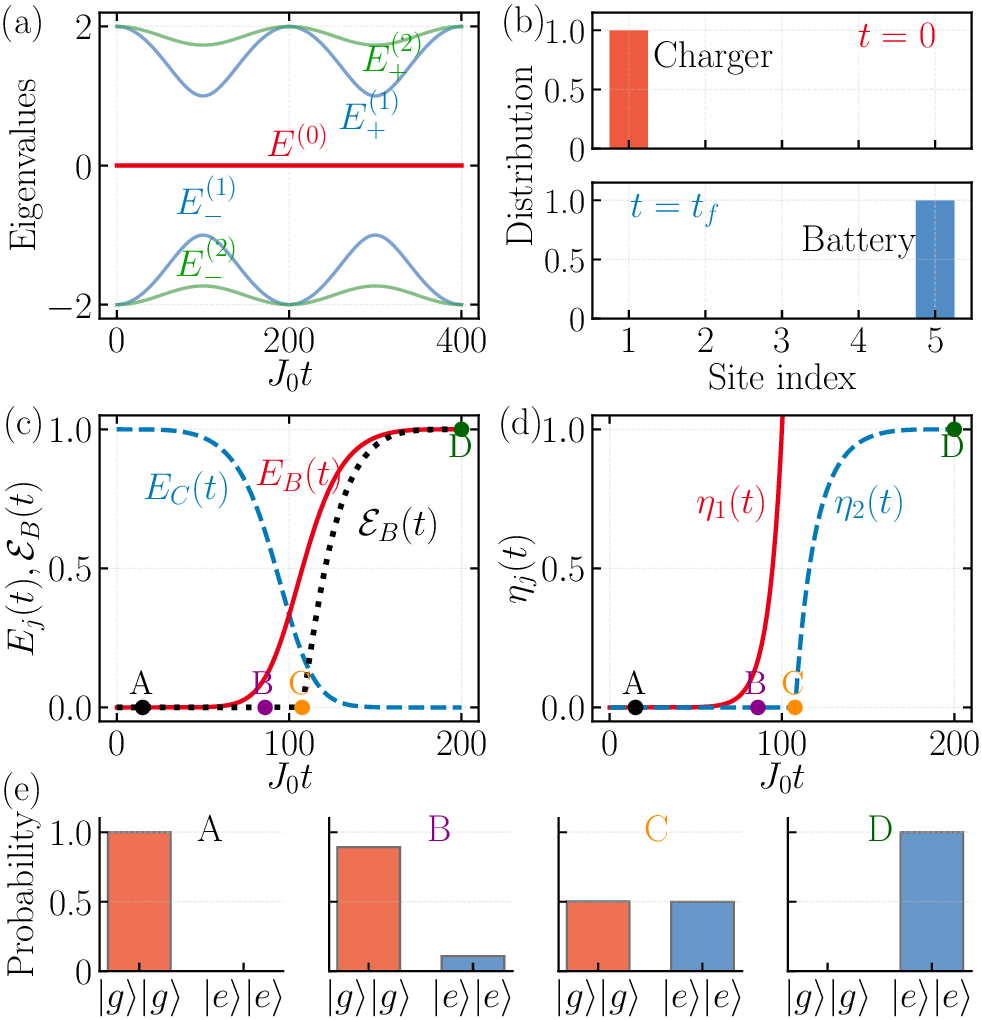}
\caption{(a) Energy spectrum versus $t$. (b) Zero-energy eigenstate at $t=0$ (upper) and $t=t_f$ (lower). (c) Time evolution of normalized charger energy $E_C(t)$ (blue dashed), battery energy $E_B(t)$ (red solid) and ergotropy $\mathcal{E}_B(t)$ (black dotted). (d) Time evolution of $\eta_1(t)$ (red solid) and $\eta_2(t)$ (blue dashed). (e) Population distributions of $\rho_B$ in $|e\rangle$ and $|g\rangle$ at the four selected time instants A, B, C, and D marked in (c) and (d). We set $t_f = 200/J_0$ and $N=5$.}
\label{fig:energy_spectrum_transfer_evolution}
\end{figure}

\textit{Topologically protected charging---}To verify the above analysis. we first adopt the commonly used cosine-shaped pulses
$f(t)=\cos(\pi t/t_f) $. For simplicity, we take $m = 2$ (i.e., $N = 5$ sites). The numerically simulated energy spectrum is presented in Fig.~\ref{fig:energy_spectrum_transfer_evolution}(a). 
It is observed that the eigenenergy $E^{(0)}$ remains zero throughout the evolution. As shown in Fig.~\ref{fig:energy_spectrum_transfer_evolution}(b), the corresponding zero-energy eigenstate is the left edge state $|L\rangle = |e,g,g,g,g\rangle$ at $t = 0$ and the right edge state $|R\rangle = |g,g,g,g,e\rangle$ at $t = t_f$, which is consistent with the above analysis. To investigate the dynamical energy transfer from the charger to the battery through the topological passage, we introduce the energy of charger $E_C$ and battery $E_B$ as
\begin{align}
E_C(t) &= \mathrm{Tr}\left[\hat{H}_C \rho_C(t)\right],\quad
E_B(t) = \mathrm{Tr}\left[\hat{H}_B \rho_B(t)\right],
\label{eq:EB}
\end{align}
where $\hat{H}_C = \omega_c \hat{\sigma}_{a_1}^{+}\hat{\sigma}_{a_1}^{-}$ and $\hat{H}_B = \omega_b \hat{\sigma}_{a_{m+1}}^{+}\hat{\sigma}_{a_{m+1}}^{-}$ are the local Hamiltonians of the charger and battery qubits, respectively, with their free frequencies $\omega_c=\omega_b$. The reduced density matrix of the charger, $\rho_C(t)$, is obtained by tracing the full density matrix $\rho(t)$ over all degrees of freedom except the charger. Similarly, the reduced density matrix of the battery, $\rho_B(t)$, is obtained by tracing $\rho(t)$ over all degrees of freedom except the battery. The density matrix $\rho(t)=|\psi(t)\rangle\langle\psi(t)|$ of the whole system can be obtained by simulating the Schr\"odinger equation  $i\partial_t |\psi(t)\rangle = H |\psi(t)\rangle$ using the standard fourth-order Runge-Kutta method. While $E_B$ denotes the stored energy of the battery, its extractability is quantified by the ergotropy, i.e., the maximum extractable work, defined as
\begin{equation}
\mathcal{E}_B(t) = E_B(t) - \min_{U} \mathrm{Tr}\left[\hat{H}_B U \rho_B(t) U^{\dagger}\right],
\label{eq:Ergdef}
\end{equation}
where the minimization is taken over all unitary operations $U$ acting locally on the battery. Explicitly, the ergotropy can be computed by
\begin{equation}
\mathcal{E}_B(t) = E_B(t) - \sum_{j} r_j \epsilon_j,
\label{eq:Erg}
\end{equation}
where $r_j$ are the eigenvalues of $\rho_B(t)$ arranged in descending order, and $\epsilon_j$ are the eigenvalues of $\hat{H}_B$ arranged in ascending order.

\begin{figure*}[t]
\centering
\includegraphics[width=0.96\textwidth]{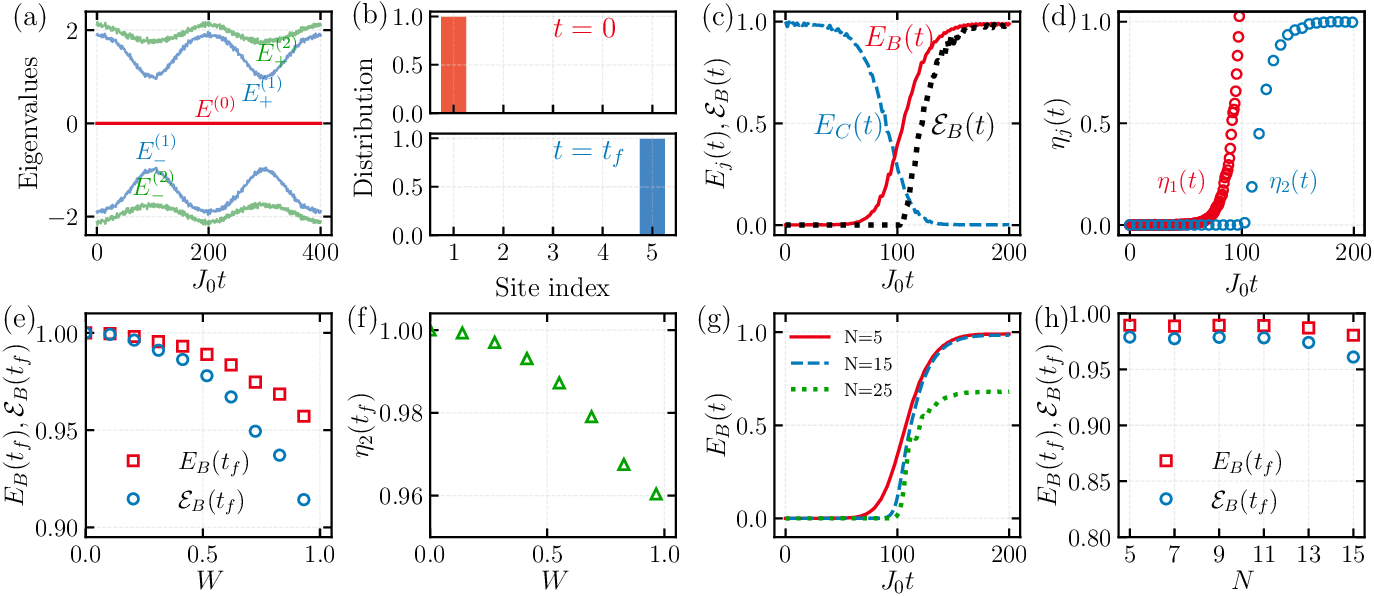}
\caption{(a) Energy spectrum with disorder. (b) Zero-energy eigenstate distributions at $t=0$ (upper) and $t=t_f$ (lower). (c) Evolution of the normalized $E_C(t)$ (blue dashed), $E_B(t)$ (red solid), and $\mathcal{E}_B(t)$ (black dotted) with disorder. (d) Evolution of efficiency $\eta_1(t)$ (red circles) and extraction efficiency $\eta_2(t)$ (blue circles). (e) Normalized $E_B(t_f)$ (red squares) and $\mathcal{E}_B(t_f)$ (blue circles) versus $W$. (f) Extraction efficiency $\eta_2(t_f)$ versus $W$. (g) Evolution of normalized $E_B(t)$ for $N = 5, 15, 25$. (h) Normalized $E_B(t_f)$ (red squares) and $\mathcal{E}_B(t_f)$ (blue circles) versus $N$. All results are averaged over $20$ disorder realizations, except for the data in (e) and (h), where we use 100 realizations to ensure convergence for larger $W$ and $N$. We set $W = 0.5$ in (a)-(d) and (g)-(h). Other parameters are the same as in Fig.~\ref{fig:energy_spectrum_transfer_evolution}.}
\label{fig:ergotropy_ABCD}
\end{figure*}

In the following numerical simulations, all energies are normalized by \(\omega_b\). For notational simplicity, we write \(E_j(t)\) and $\mathcal{E}_B(t)$ to denote the normalized quantity \(E_j(t)/\omega_b\) and $\mathcal{E}_B(t)/\omega_b$ in the figures. In Fig.~\ref{fig:energy_spectrum_transfer_evolution}(c), the evolutions of normalized $E_B(t)$, $E_C(t)$ and $\mathcal{E}_B(t)$ are simulated numerically. It is observed that, as the evolution proceeds, the energy of the charger decreases while that of the battery increases. At the target time $t = t_f$, all energy initially stored in the charger is completely transferred to the battery. Unlike the evolution of $E_B(t)$, $\mathcal{E}_B(t)$ stays zero initially, reaches a critical point C, and then grows with time in a manner similar to the stored energy. To further characterize the charging process and extraction efficiency, we define  the ratio $\eta_1(t) = \frac{E_B(t)}{E_C(t)}$ and $
\eta_2(t) = \frac{\mathcal{E}_B(t)}{E_B(t)}$. As shown in Fig.~\ref{fig:energy_spectrum_transfer_evolution}(d), $E_B(t)$ and $E_C(t)$ become equal at $t = t_f/2$, and for $t > t_f/2$, the battery energy exceeds that of the charger. At the target time $t = t_f$, $\eta_1(t_f)$ approaches infinity. Moreover, $E_B(t)$, $E_C(t)$, and $\eta_1(t)$ are all monotonic, indicating that energy is continuously charging to the battery without any backflow.
$\eta_2(t)$ follows a trend similar to $\mathcal{E}_B(t)$, with $\eta_2(t) = 0$ before a critical point C and increasing thereafter. To explain why the stored energy is not extractable before the critical point, we perform quantum state tomography on $\rho_B$. The off-diagonal elements of $\rho_B$ are found to be zero throughout the evolution, so that $\rho_B = \rho_{ee}|e\rangle\langle e| + \rho_{gg}|g\rangle\langle g|$. Figure~\ref{fig:energy_spectrum_transfer_evolution}(e) shows the populations $\rho_{ee}$ and $\rho_{gg}$ at the four time points A, B, C, and D indicated in Figs.~\ref{fig:energy_spectrum_transfer_evolution}(c) and~(d). At the critical point C, we have $\rho_{ee} = \rho_{gg}$. Before the critical point, as seen at points A and B, $\rho_{ee} < \rho_{gg}$, and no work can be extracted. After the critical point, $\rho_{ee} > \rho_{gg}$, which is precisely the population inversion condition for incoherent ergotropy, and thus the stored energy becomes extractable. Nevertheless, at point D ($t = t_f$), $\rho_{gg} = 0$, the state $\rho_B$ is pure, $\eta_2(t)=1$, confirming that the stored energy becomes completely extractable by the end of the protocol.

To verify the topologically protected nature of the charging process, we introduce disorder into the system. In our model, the primary tunable parameter is the coupling $J_i(t)$. To simulate control imperfections, we introduce disorder into the couplings $J_i(t)$ as $J_i \to J_i + \delta J_i$, where $\delta J_i = W \delta$, with $W$ the disorder strength and $\delta \in [-0.5, 0.5]$ a Gaussian random variable. In Figs.~\ref{fig:ergotropy_ABCD}(a) and~(b), we numerically simulate the energy spectrum of the disordered system with $W=0.5$ and the corresponding probability distributions of the zero-energy eigenstate at the initial and target times, respectively. It is observed that, the nonzero-energy branches of the spectrum become non-smooth due to the disorder, while the zero-energy branch remains smooth. Correspondingly, the associated eigenstate is localized at the left edge at $t = 0$ and at the right edge at $t = t_f$, which is identical to the ideal-case distribution shown in Fig.~\ref{fig:energy_spectrum_transfer_evolution}(b). This demonstrates that the edge states are immune to disorder, manifesting the topological protection. Dynamically, as shown in Fig.~\ref{fig:ergotropy_ABCD}(c), the charger energy is fully transferred to the battery even with disorder. The ergotropy evolution is nearly identical to the ideal case, indicating that the disorder with $W=0.5$ has negligible effect. It is precisely this topologically protected edge state that enables a disorder-robust quantum battery protocol. Correspondingly, $\eta_1(t_f)$ diverges, as seen in Fig.~\ref{fig:ergotropy_ABCD}(d), and the extraction efficiency $\eta_2(t_f)$ remains close to $100\%$, demonstrating resilience against control imperfections.

We next discuss the effects of stronger disorder and larger system sizes on the battery performance. In Fig.~\ref{fig:ergotropy_ABCD}(e), we simulate the normalized stored energy $E_B(t_f)$ and the ergotropy $\mathcal{E}_B(t_f)$ as a function of $W$. It is observed that, within a certain range of $W$, the presence of disorder has little effect on either quantity. However, when $W$ becomes too large, both $E_B(t_f)$ and $\mathcal{E}_B(t_f)$ decrease with increasing $W$, with the extractable work being more severely affected. This indicates that disorder does alter the quantum state $\rho_B$ of the battery. Nevertheless, even for $W = 1$, we have $E_B(t_f)/\omega_b$ larger than $0.95$ and $\mathcal{E}_B(t_f)/\omega_b$ larger than $0.9$, corresponding to reductions of only $5\%$ and $10\%$ relative to the ideal case, respectively. This demonstrates that even strong disorder does not significantly degrade the battery performance. Moreover, although $\mathcal{E}_B(t_f)$ decreases more rapidly with $W$, the extraction efficiency $\eta_2(t_f)$ of the stored energy remains above $96\%$ for $W = 1$, as shown in Fig.~\ref{fig:ergotropy_ABCD}(f), further confirming the robustness of the battery performance against disorder. In Fig.~\ref{fig:ergotropy_ABCD}(g), we fix $W = 0.5$ and simulate the evolution of the normalized $E_B(t)$ for different system sizes $N = 5$, $15$, and $25$. The results show that the overall evolution trends are consistent across different $N$, but the time at which $E_B(t)$ begins to rise significantly increases with $N$. Moreover, for too large $N$, complete energy transfer from the charger to the battery at $t = t_f$ is no longer achieved. To further investigate the finite-size effect on the stored energy and extractable work, we plot in Fig.~\ref{fig:ergotropy_ABCD}(h) the normalized $E_B(t_f)$ and $\mathcal{E}_B(t_f)$ as a function of $N$. It is found that both $E_B(t_f)$ and $\mathcal{E}_B(t_f)$ decrease with increasing $N$, with the latter exhibiting a more rapid decline. Nevertheless, both quantities remain above $0.95$ for $N \le 15$, confirming that our scheme tolerates a moderately large topological chain length. For larger $N$, however, achieving high performance requires more precise control over the couplings to reduce the disorder strength.

\textit{Accelerating topologically protected charging---}The charging speed of the battery is another important figure of merit. From the above analysis, it seems that simply reducing the target time $t_f$ in the control fields could accelerate the topologically protected charging. However, a further investigation shows that a smaller $t_f$ violates the adiabatic condition, leading to a degradation of both the stored energy and the ergotropy (see the Supplemental Material). To accelerate the topologically protected charging process, we choose an alternative control field $f(t) = \tan(\pi t/T+\alpha)/\tan(\alpha)$, with $T = \pi t_f/[2(\pi-\alpha)]$ and $\alpha = \pi/2 + \mu$, where $\mu$ modulates the flatness of the pulse shape.

\begin{figure}[t]
\centering
\includegraphics[width=0.48\textwidth]{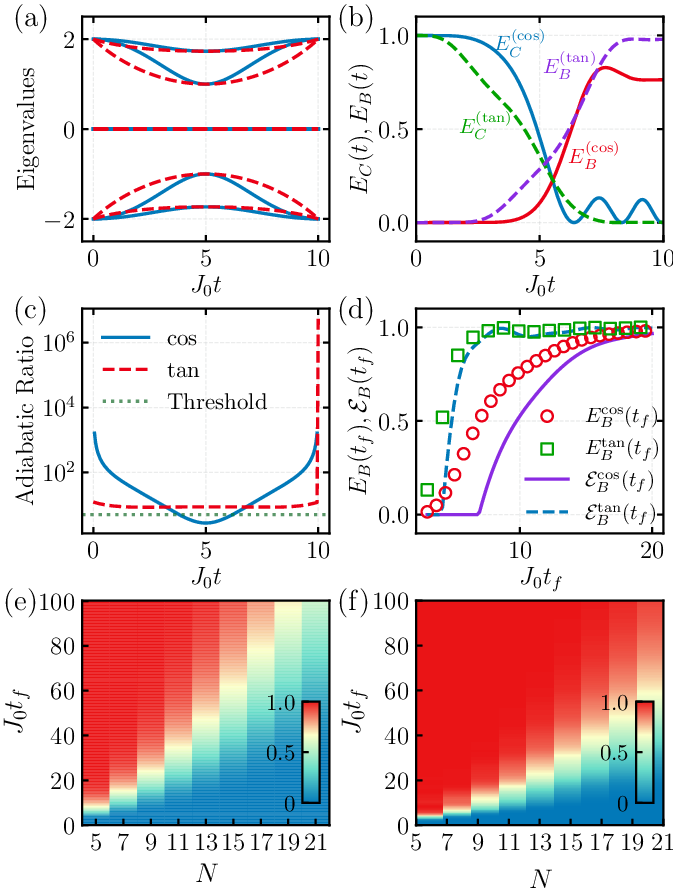}
\caption{(a) Energy spectra for cosine (blue solid) and tangent (red dashed) controls. (b) Normalized $E_C(t)$ and $E_B(t)$ under cosine (solid) and tangent (dashed) controls. (c) Adiabatic ratio $C(t)$ for cosine (blue solid) and tangent (red dashed) controls; the green dotted line marks the threshold. (d) Normalized $E_B(t_f)$ and $\mathcal{E}_B(t_f)$ versus $t_f$ for both controls. (e) Density plot of normalized $E_B(t_f)$ versus $N$ and $t_f$ for cosine control. (f) Density plot of normalized $E_B(t_f)$ versus $N$ and $t_f$ for tangent control. We set $t_f=10/J_0$ in (a)-(c) and $\mu=0.8$ throughout.}
\label{fig:noise_spectrum_transfer_evolution}
\end{figure}

For comparison, we numerically simulate the energy spectra for both the cosine-shaped and tangent-shaped pulses in Fig.~\ref{fig:noise_spectrum_transfer_evolution}(a), where the blue solid curve corresponds to the cosine pulse and the red dashed curve to the tangent pulse. Both spectra exhibit a zero-energy eigenvalue that remains constant in time. However, the nonzero eigenvalues show significant differences between the two cases. For the cosine-shaped pulse, the nonzero eigenvalues vary slowly near the two ends. In contrast, for the tangent-shaped pulse, the nonzero eigenvalues vary rapidly near the two ends and slowly in the middle. Dynamically, as shown in Fig.~~\ref{fig:noise_spectrum_transfer_evolution}(b), the charging process under the tangent pulse exhibits fast evolution at early times and slow evolution in the intermediate regime, compared with the cosine case. At $t = t_f$, the tangent pulse yields significantly better charging performance than the cosine pulse.

To explain this observation, we recall the adiabatic condition derived in Supplemental Material and define the adiabatic ratio
\begin{equation}
C = \min_{l \neq m+1} \frac{\left| E_{m+1} - E_l \right|}{\left| \left( S^{-1} \dot{S} \right)_{m+1, l} \right|}.
\label{eq:adiabatic_ratio}
\end{equation}
To satisfy the adiabatic condition, $C\gg1$ is required. Based on our numerical results, we set a threshold of $5$, with larger values indicating better adiabaticity. In Fig.~\ref{fig:noise_spectrum_transfer_evolution}(c), we plot the time evolution of $C$ for both the cosine-shaped and tangent-shaped pulses. For the cosine pulse, $C$ is large near the two ends where the variation in Fig.~\ref{fig:noise_spectrum_transfer_evolution}(a) is slow, but drops below the threshold of 5 in the middle of the evolution, indicating that the adiabatic condition is violated and other eigenstates become excited. As a result, at the target time, a portion of the energy remains in other bulk states. For the tangent pulse, by contrast, $C$ remains above the threshold throughout the evolution, ensuring that the adiabatic condition is well satisfied and that almost no other bulk states are excited. Consequently, the charger energy can be transferred to the battery almost completely via the topologically protected channel. This suggests that, in designing the control fields, a profile that varies rapidly at the initial and final stages, while slowly in the middle, is more favorable for satisfying the adiabatic condition and thus accelerating charging. As for the optimal control field, it could be further explored using reinforcement learning methods~\cite{h38x-bh1j}, but this is beyond the scope of the present paper. Nevertheless, the tangent-shaped pulse can accelerate the charging process compared with the cosine-shaped one. To quantify the speedup, we simulate in Fig.~\ref{fig:noise_spectrum_transfer_evolution}(d) the normalized stored energy and the extractable work as functions of $t_f$ for both pulses. For the cosine pulse, a target time of at least $t_f \approx 20/J_0$ is required to achieve ideal charging, whereas for the tangent pulse, $t_f \approx 7/J_0$ suffices. Thus, the charging process is significantly accelerated. To further compare the acceleration effect for different chain lengths $N$, we show in Figs.~\ref{fig:noise_spectrum_transfer_evolution}(e) and (f) the normalized $E_B(t_f)$ as a function of $t_f$ and $N$ for the cosine pulse and the tangent pulse, respectively. Both panels reveal that larger $N$ requires slower control-field variation, and hence a longer charging time, for optimal performance. This is because a larger $N$ corresponds to a higher-dimensional Hilbert space, which introduces more bulk states that may couple to the edge state. To suppress such crosstalk, the control fields must be varied more gradually to satisfy the adiabatic condition for larger $N$. However, for the same $N$, the tangent pulse requires a smaller $t_f$. Therefore, the acceleration effect is universal with respect to the system size $N$.

\textit{Conclusion---}In conclusion, we have demonstrated a robust and efficient long-distance topological charging protocol. We first showed that, under time-dependent control, the topologically protected zero-energy edge state enables near-perfect adiabatic remote energy transfer from the charger to the battery without energy backflow. We then investigated the robustness of this topological charging channel in the presence of disorder and found that the charging performance remains highly stable against such perturbations. Furthermore, by analyzing the dependence on the number of qubits, we demonstrated the scalability of the protocol and found that topological protection effectively suppresses performance degradation over a moderate range of system sizes. Importantly, an analysis of the ergotropy confirms that the energy transferred to the battery is almost fully extractable, highlighting the high charging efficiency of the protocol. Finally, we compared cosine- and tangent-type time-dependent control fields and found that the latter substantially accelerates the charging dynamics, resulting in significant enhancements in both the stored energy and ergotropy. Overall, our results establish a flexible, robust, and scalable strategy for topological quantum battery charging without energy backflow, and provide a promising framework for optimizing charging dynamics through time-dependent control.

This work was supported by the National Natural Science Foundation of China (Grant No. 12104141), the Natural Science Foundation Innovation-Development Joint  Program of Hubei Province (Grant No. 2026AFC0003), the Key Project of the Scientific Research Program of the Department of Education of Hubei Province (Grant No. D20252503).

\bibliographystyle{apsrev4-1}
\bibliography{topobattery}

\clearpage
\pagestyle{empty}
\onecolumngrid
\vspace*{10pt}
\renewcommand{\theequation}{S\arabic{equation}}
\setcounter{equation}{0}
\begin{bibunit}[apsrev4-2]
\begin{center}
	\large \textbf{Supplemental Material for “Pulse-Controlled Topologically Protected Quantum Batteries”}
\end{center}

\section{Eigenvalues and Eigenstates of SSH model}
In this section, we write the model in the matrix form, and present the corresponding Eigenvalues and Eigenstates. For the Hamiltonian used in the maintext
\begin{align}
\hat{H}(t)=\sum_{x=1}^{m}\left[J_1(t)\hat{\sigma}_{a_x}^{+}\hat{\sigma}_{b_x}^{-}+J_2(t)\hat{\sigma}_{b_x}^{+}\hat{\sigma}_{a_{x+1}}^{-}+\mathrm{H.c.}\right],
\label{eq:H}
\end{align}
the basis vectors $\{|e,g,g,\ldots\rangle, |g,e,g,\ldots\rangle, \ldots |g,\ldots,g,e\rangle\}$ form a complete set in single excitation. In this basis, the above Hamiltonian takes the matrix form $H_M(t)$ as
\begin{align}
H_{M}(t)=\left(\begin{array}{cccccc}
0 & J_1(t) & & & & \\
J_1(t) & 0 & J_2(t) & & & \\
& J_2(t) & 0 & J_1(t) & & \\
& & \ddots & \ddots & \ddots & \\
& & & J_1(t) & 0 & J_2(t) \\
& & & & J_2(t) & 0
\end{array}\right)_{N \times N}.
\label{eq:HM}
\end{align}
For analytical convenience, we set $N=2m+1$. The eigenvalues of $H_M$ are then given by
\begin{align} 
E^{(0)} &= 0, \label{eq:E0} \\
E_{\pm}^{(k)} &= \pm J_2 \sqrt{1+\nu^2+2\nu\cos\left(\frac{k\pi}{m+1}\right)},
\end{align}
with $k=1,2,\ldots,m$ and $\nu = J_1/J_2$. Correspondingly, the eigenstates are
\begin{align} \label{eq:eigenstate0}
|\psi\rangle^{(0)}=\frac{1}{\sqrt{\eta_0}}\left[(-\nu)^0, 0,(-\nu)^1, 0, \ldots, (-\nu)^{m}\right]^T,
\end{align}
and
\begin{align} \label{eq:eigenstate1}
|\psi\rangle_{ \pm}^{(k)}=&& \frac{1}{\sqrt{\eta_{ \pm}^{(k)}}}\left[\frac{1}{\nu} \sin \left(\frac{0 k \pi}{m+1}\right)+\sin \left(\frac{k \pi}{m+1}\right) \right., \left. \frac{\lambda_{ \pm}^{(k)}}{J_2 \nu} \sin \left(\frac{k \pi}{m+1}\right), \frac{1}{\nu} \sin \left(\frac{k \pi}{m+1}\right)+\sin \left(\frac{2 k \pi}{m+1}\right),\right. \notag\\
&&\left.\frac{\lambda_{ \pm}^{(k)}}{J_2 \nu} \sin \left(\frac{2 k \pi}{m+1}\right), \ldots,  \frac{\lambda_{ \pm}^{(k)}}{J_2 \nu} \sin \left(\frac{m k \pi}{m+1}\right),\right. \left.\frac{1}{\nu} \sin \left(\frac{m k \pi}{m+1}\right)+\sin \left(\frac{(m+1) k \pi}{m+1}\right)\right]^T,
\end{align}
where the normalized coefficients  $\eta_0=\sum_{n^{\prime}=0}^m \nu^{2 n^{\prime}}$ and $\eta_{ \pm}^{(k)}=(m+1)(\lambda_{ \pm}^{(k)})^2/(J_2^2 \nu^2)$.

\section{Adiabatic Condition and Its Validation}\label{sec:adiabatic}
In this section, we derive why simply reducing the target time $t_f$ in the control fields to accelerate the charging of the quantum battery is not feasible. In fact, the analysis of the energy spectrum is static in nature. To derive the adiabatic condition, we consider the time-dependent Schrödinger equation in the matrix representation,
\begin{align}
i\frac{\partial}{\partial t} |\psi(t)\rangle = H_M(t) |\psi(t)\rangle,
\end{align}
where 
\begin{equation}
|\psi(t)\rangle = c_1|e,g,g\cdots\rangle + c_2|g,e,g\cdots\rangle + \cdots =
\begin{pmatrix}
c_1 \\ 
c_2 \\
\vdots
\end{pmatrix}.
\end{equation}
We next analyze the dynamics of the system in the eigenstate representation. To diagonalize the Hamiltonian, we introduce a similarity transformation
\begin{equation}
H_{\Lambda} = S^{-1} H_M S = 
\begin{pmatrix}
E_{-}^{(m)} & & & & & \\
& \ddots & & & & \\
& & E^{(0)} & & & \\
& & & \ddots & & \\
& & & & E_{+}^{(m)} & \\
\end{pmatrix}_{N \times N}.
\end{equation}
Here the matrix $S = (|\psi\rangle_{-}^{(m)}, \ldots, |\psi\rangle^{(0)},\ldots, |\psi\rangle_{+}^{(m)})$ is composed of the corresponding eigenstates expressed in Eqs.~(\ref{eq:eigenstate0}) and (\ref{eq:eigenstate1}). Applying this transformation to the time-dependent Schr\"odinger equation and using the relation $\dot{S}^{-1} S = -S^{-1} \dot{S}$, we obtain
\begin{equation}
i\frac{\partial}{\partial t} |\psi_{\Lambda}\rangle = H_{\Lambda} |\psi_{\Lambda}\rangle - i S^{-1} \dot{S} |\psi_{\Lambda}\rangle,
\label{eq:adiabatic_basis}
\end{equation}
where 
\begin{equation}
|\psi_{\Lambda}(t)\rangle = S^{-1} |\psi(t)\rangle =
\begin{pmatrix}
d_{-}^{(m)} \\ 
\vdots\\
d^{(0)} \\
\vdots\\
d_{+}^{(m)}
\end{pmatrix}.
\end{equation}
is the quantum state in the instantaneous eigenbasis with $d_{\pm}^{(k)}=[\langle \psi | \psi\rangle_{\pm}^{(k)}]^*$ and $d^{(0)}=[\langle \psi | \psi\rangle^{(0)}]^*$. Therefore, if the second term $- i S^{-1} \dot{S} |\psi_{\Lambda}\rangle$ in Eq.~\eqref{eq:adiabatic_basis} is neglected, the system evolves within each eigenstate independently by
\begin{equation}
i\frac{\partial}{\partial t}
\begin{pmatrix}
d_{-}^{(m)} \\ 
\vdots\\
d^{(0)} \\
\vdots\\
d_{+}^{(m)}
\end{pmatrix}=\begin{pmatrix}
E_{-}^{(m)} & & & & & \\
& \ddots & & & & \\
& & E^{(0)} & & & \\
& & & \ddots & & \\
& & & & E_{+}^{(m)} & \\
\end{pmatrix}\begin{pmatrix}
d_{-}^{(m)} \\ 
\vdots\\
d^{(0)} \\
\vdots\\
d_{+}^{(m)}
\end{pmatrix}.
\end{equation}
Hence, if the system is initially prepared in the zero-energy eigenstate $|\psi\rangle^{(0)}=\frac{1}{\sqrt{\eta_0}}\left[(-\nu)^0, 0,(-\nu)^1, 0, \ldots, (-\nu)^{m}\right]^T$, it will remain in that eigenstate throughout the evolution. In this case, the dynamical results are consistent with the energy spectrum analysis: by adiabatically varying the parameter $\nu$ from $|\nu|\ll 1$ to $|\nu| \gg 1$, the zero-energy eigenstate evolves from the left edge state to the right edge state, thereby realizing topologically protected charging. However, the presence of the second term induces couplings between different eigenstates. To make these couplings weak, the system must be kept far from resonance in the eigenstate representation, i.e.,
\begin{equation} \label{Adiabatic_cond}
\left| \left( S^{-1} \dot{S} \right)_{kl} \right| \ll \left| E_k - E_l \right|, \quad (l \neq k),
\end{equation}
where $E_k$ is the $k$-th eigenvalue. For the zero-energy state, we have $k = m+1$, i.e., $E_{m+1} = E^{(0)}$. More specifically, to avoid crosstalk with the zero-energy eigenstate, the adiabatic condition can be written as
\begin{equation} \label{Adiabatic_cond_zero}
\left| \left( S^{-1} \dot{S} \right)_{m+1, l} \right| \ll \left| E_{m+1} - E_l \right|, \quad (l \neq m+1).
\end{equation}
For our proposal
\begin{align}
J_1(t) &= J_0\left[1 - f(t) \right], \quad
J_2(t) = -J_0\left[1 + f(t)\right],
\label{eq:J_pulses}
\end{align}
with $f(t)=\cos(\pi t/t_f) $ for example. When $t_f$ becomes too small, the control fields $J_1(t)$ and $J_2(t)$ in Eq.~\eqref{eq:J_pulses} vary too rapidly, resulting in a large $\dot{S}$. As a consequence, the adiabatic condition above is no longer satisfied. Therefore to reduce $t_f$ to accelerate the charging process will degrade both the stored energy and the extractable work. 

\begin{figure*}[t]
\centering
\includegraphics[width=0.8\textwidth]{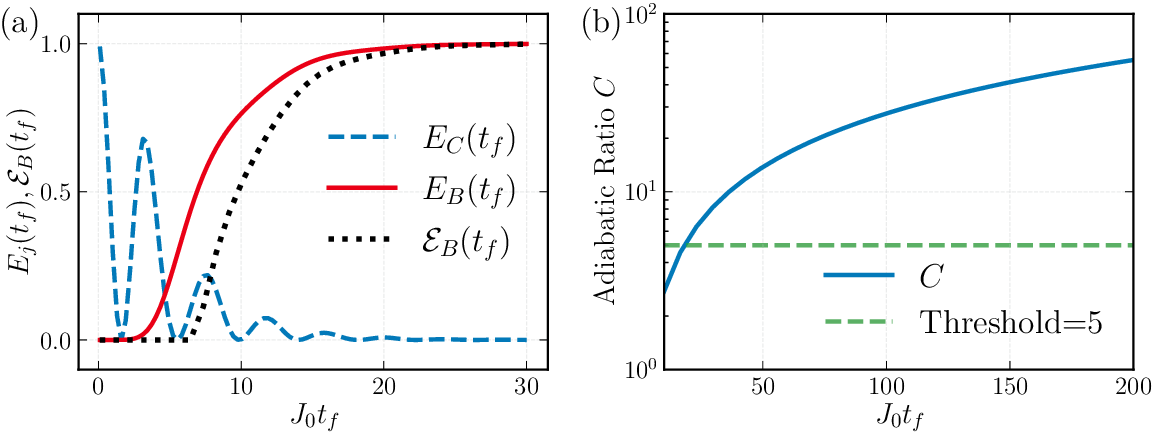}
\caption{(a) Normalized $E_C(t_f)$ (blue dashed), $E_B(t_f)$ (red solid), and $\mathcal{E}_B(t_f)$ (black dotted) versus $t_f$. (b) Adiabatic ratio $C$ as a function of $t_f$. We set $N=5$}
\label{fig:energy_ergotropy_tf}
\end{figure*}

To verify this analysis, we take cosine-shaped pulses
$f(t)=\cos(\pi t/t_f) $ as a example. In Fig.~\ref{fig:energy_ergotropy_tf}(a), we simulate the normalized charger energy $E_C(t_f)$ , battery energy $E_B(t_f)$, and ergotropy $\mathcal{E}_B(t_f)$ as a function of $t_f$. It is found that a moderate reduction of $t_f$ indeed accelerates the charging process without degrading performance. However, when $t_f$ becomes too small, the battery energy $E_B(t_f)$ no longer reaches unity, and the charger energy $E_C(t_f)$ no longer vanishes, indicating that complete energy transfer from the charger to the battery is no longer achieved. Consequently, an excessively small $t_f$ also reduces the extractable work. To explain this, we define the adiabatic ratio
\begin{equation}
C = \min_{l \neq m+1} \frac{\left| E_{m+1} - E_l \right|}{\left| \left( S^{-1} \dot{S} \right)_{m+1, l} \right|},
\label{eq:adiabatic_ratio}
\end{equation}
based on the condition in Eq.~(\ref{Adiabatic_cond_zero}). The adiabatic condition requires $C \gg 1$. Based on our numerical results, we adopt a threshold of $5$, with larger $C$ indicating better adiabaticity. For each $t_f$, we further tighten the definition of the adiabatic ratio as
\begin{equation}
C = \min_{t} \min_{l \neq m+1} \frac{\left| E_{m+1} - E_l \right|}{\left| \left( S^{-1} \dot{S} \right)_{m+1, l} \right|},
\label{eq:adiabatic_ratio2}
\end{equation}
where the minimization over $t$ is taken over the entire evolution from $0$ to $t_f$.

As shown in Fig.~\ref{fig:energy_ergotropy_tf}(b), we investigate adiabatic ratio $C$ expressed in Eq.~(\ref{eq:adiabatic_ratio2}) as a function of the target time $t_f$. It is observed that as the target time $t_f$ increases, the control fields vary more slowly and the adiabatic ratio $C$ increases accordingly, indicating that a larger $t_f$ better satisfies the adiabatic condition. For smaller $t_f$, however, the control fields vary too rapidly, causing $C$ to fall below the threshold. Consequently, the adiabatic condition is violated, and other nonzero eigenstates (i.e. bulk states) become populated, trapping part of the energy. This explains why, for small $t_f$ in Fig.~\ref{fig:energy_ergotropy_tf}(a), the charger energy cannot be fully transferred to the battery. Moreover, since a larger $N$ corresponds to a higher-dimensional Hilbert space, which introduces more eigenstates that may couple to the zero-energy eigenstate. Thus larger system sizes $N$ require a longer target time $t_f$ to maintain a high battery stord energy as shown in the maintext.

\section{Robustness of Accelerated Topologically Protected Charging}\label{Topo_charge_optmi}

\begin{figure*}
\centering
\includegraphics[width=0.8\textwidth]{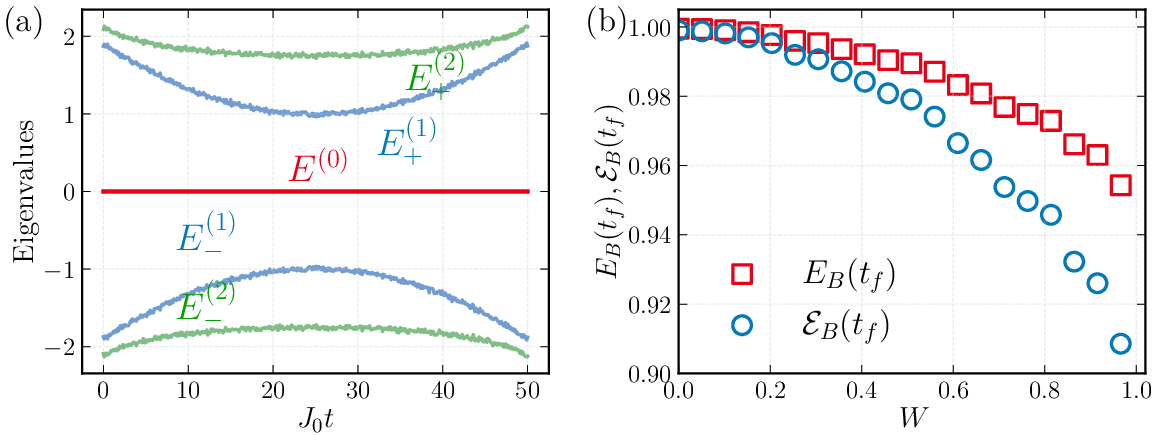}
\caption{(a) Energy spectrum for tan-type control with $t_f=50/J_0$ and disorder strength $W=0.5$. (b) Normalized $E_B(t_f)$ (red squares) and normalized $\mathcal{E}_B(t_f)$ (blue circles) versus $W$ for $t_f=50/J_0$. Other parameters are as in Fig.~\ref{fig:energy_ergotropy_tf}.}
\label{fig:tan_robust}
\end{figure*}

In this section, we discuss the robustness for the accelerated topologically protected charging process using control field 
\begin{eqnarray}
f(t) &=& \frac{\tan\left(\pi t /T+\alpha\right)}{\tan(\alpha)},
\end{eqnarray}
where $T = \dfrac{\pi t_f}{2(\pi - \alpha)}$, and $\alpha=\frac{\pi}{2}+\mu$ with $\mu$ a constant. When the control fields are subject to noise, the coupling strength is modified as $J_i \to J_i + \delta J_i$, with $\delta J_i = W \delta$. As shown in Fig.~\ref{fig:tan_robust}(a), for a noise amplitude $W = 0.5$, the nonzero eigenvalues of the spectrum become non-smooth as a function of $t$, while the zero eigenenergy $E^{(0)}$ remains exactly zero throughout the evolution, unaffected by the disorder.

To investigate the topological protection under tangent pulse coupling, we plot in Fig.~\ref{fig:tan_robust}(b) the normalized $E_B(t_f)$ and $\mathcal{E}_B(t_f)$ versus $W$. Even at $W \approx 1$, the losses in stored energy and extractable work are below $5\%$ and $10\%$, respectively, confirming that the tangent pulse preserves topological protection. The robustness is comparable to that of the cosine pulse case shown in Fig.3(e) in the maintext, but with $t_f = 50$, i.e., one quarter of the value used for the cosine pulse. This indicates that the tangent pulse achieves both speedup and disorder resilience similar to the cosine pulse.

\end{bibunit}

\end{document}